\documentclass{sig-alternate-05-2015}
\usepackage{xcolor}
\usepackage{subcaption}
\usepackage{balance}
\usepackage{pdfpages}
\usepackage{graphicx}
\usepackage{booktabs}
\usepackage{subcaption}

\begin{document}

% Copyright
%\setcopyright{acmcopyright}
%\setcopyright{acmlicensed}
%\setcopyright{rightsretained}
%\setcopyright{usgov}
%\setcopyright{usgovmixed}
%\setcopyright{cagov}
%\setcopyright{cagovmixed}

% DOI
%\doi{10.475/123_4}

% ISBN
%\isbn{123-4567-24-567/08/06}

%Conference
%\conferenceinfo{PLDI '13}{June 16--19, 2013, Seattle, WA, USA}

%\acmPrice{\$15.00}

%
% --- Author Metadata here ---
%\conferenceinfo{WOODSTOCK}{'97 El Paso, Texas USA}
%\CopyrightYear{2007} % Allows default copyright year (20XX) to be over-ridden - IF NEED BE.
%\crdata{0-12345-67-8/90/01}  % Allows default copyright data (0-89791-88-6/97/05) to be over-ridden - IF NEED BE.
% --- End of Author Metadata ---
\title{Investigating Gender Bias in Touch Biometrics}
%\subtitle{[Extended Abstract]
%\titlenote{A full version of this paper is available as
%\textit{Author's Guide to Preparing ACM SIG Proceedings Using
%\LaTeX$2_\epsilon$\ and BibTeX} at
% For aesthetic reasons, we recommend 'three authors at a time
% i.e. three' name/affiliation blocks' be placed beneath the title.
% More than six makes the first page appear very cluttered indeed.

\numberofauthors{3} %  in this sample file, there are a *total*
% of EIGHT authors. SIX appear on the 'first page' (for formatting
% \author{
% % You can go ahead and credit any number of authors here,
% % e.g. one' row of three' or two rows (consisting of one row of three
% % and a second row of one, two or three).
% %
% % The command \alignauthor (no curly braces needed) should
% % precede each author name, affiliation/snail-mail address and
% % e-mail address. Additionally, tag each line of
% % affiliation/address with \affaddr, and tag the
% % e-mail address with \email.
% %
% % 1st. author
% \alignauthor
% Joshua Lee \\
%         \affaddr{Bucknell University, PA}\\
%         \email{jjl032@bucknell.edu}
% % 2nd. author
% \alignauthor
% Ben Khant \\
%         \affaddr{Bucknell University, PA}\\
%         \email{ak058@bucknell.edu }
% % 3rd. author
% \alignauthor
% Rajesh Kumar\\
%        \affaddr{Bucknell University, PA}\\
%        \email{rk042@bucknell.edu}
% }

\author{
Joshua Lee\thanks{Joshua Lee and Ben Khant contributed equally to this work. Accepted for presentation at The Richard Tapia Conference (Tapia 2026)}\\
Bucknell University, PA\\
jjl032@bucknell.edu
\and
Ben Khant\footnotemark[1]\\
Bucknell University, PA\\
ak058@bucknell.edu
\and
Rajesh Kumar\\
Bucknell University, PA\\
rk042@bucknell.edu
}

\maketitle

\begin{abstract}
Behavioral biometrics offer a promising approach for continuous authentication, but their fairness across demographic groups remains largely unexplored. This paper investigates gender bias in swipe-based authentication using the BBMAS (117 users) and ANTAL (71 users) datasets and evaluates XGBoost and DenseNet classifiers through False Acceptance Rate (FAR) and False Rejection Rate (FRR). XGBoost achieved authentication accuracies of 92\% and 94\% on the BBMAS and ANTAL datasets, respectively, while statistical tests (Kolmogorov-Smirnov, Mann-Whitney, and Wasserstein permutation) found no significant gender differences in authentication error rates across almost all experimental settings. These findings suggest that swipe-based authentication can achieve high accuracy while maintaining comparable performance for male and female users, supporting its potential as a fair and reliable behavioral biometric modality.

\end{abstract}
 
\section{Introduction}
User authentication is an essential component of smartphone security, particularly as mobile devices store sensitive personal, financial, and organizational information \cite{patel2016continuous,TouchSurveyTechniques}. Traditional authentication mechanisms, such as PINs, passwords, fingerprints, and facial recognition, typically verify identity only at the point of login \cite{Jain2021BiometricsTB}. Once access is granted, these mechanisms provide limited protection against session hijacking, device sharing, coercion, or unauthorized use after initial authentication. Continuous authentication based on behavioral biometrics addresses this limitation by repeatedly verifying user identity during an active session using natural interaction patterns such as swiping, typing, and device movement \cite{frank2012touchalytics,kumar2016continuous,TouchSurveyTechniques}.

Among behavioral biometric modalities, swipe gestures are particularly attractive for smartphone authentication because they are generated naturally during everyday device use and can be captured using standard touchscreen sensors without requiring additional hardware \cite{frank2012touchalytics,Li2013UnobservableRF,FairFair}. Prior work has shown that swipe-based authentication can distinguish genuine users from impostors with promising accuracy, making it suitable as a second-layer authentication mechanism \cite{frank2012touchalytics,serwadda2013verifiers,TouchSurveyTechniques}. However, high authentication accuracy alone is not sufficient for trustworthy deployment \cite{Jain2021BiometricsTB, BiasTBIOM2020}. A biometric system must also be evaluated for whether its errors are distributed equitably across demographic groups \cite{Jain2021BiometricsTB, BiasTBIOM2020}.

Fairness has become a critical concern in biometric systems because demographic disparities can lead to unequal security and usability outcomes \cite{Jain2021BiometricsTB}. For example, a group with a higher False Rejection Rate (FRR) may experience more frequent lockouts, while a group with a higher False Acceptance Rate (FAR) may receive weaker protection against impostors. Prior studies have documented demographic bias in biometric systems, particularly in face recognition, motivating fairness evaluation for emerging biometric modalities before large-scale deployment \cite{buolamwini2018gender,BiasTBIOM2020,Jain2021BiometricsTB}. Despite extensive research on the performance and robustness of touch-based authentication, its fairness across demographic groups remains less understood, partly because most publicly available touch gesture datasets do not include demographic attributes \cite{TouchSurveyTechniques,FairFair}.

This paper investigates whether swipe-based authentication exhibits gender-based disparities in authentication error rates. We use two publicly available datasets containing gender labels, BBMAS \cite{belman2019insights} and ANTAL \cite{antal2016biometric}, and evaluate two authentication models: XGBoost, representing a classical machine learning classifier, and DenseNet, representing a deep learning architecture. We assess authentication performance using False Acceptance Rate (FAR) and False Rejection Rate (FRR), and we compare the corresponding error distributions across male and female users using statistical hypothesis testing.

\underline{\textit{Problem definition}} Let \( \mathcal{D} = \{(\mathbf{x}_i, y_i, g_i)\}_{i=1}^{n} \) denote a swipe-authentication dataset, where \( \mathbf{x}_i \in \mathbb{R}^{d} \) is the feature vector extracted from the \(i\)-th swipe gesture, \( y_i \in \{0,1\} \) is the authentication label, and \( g_i \in \mathcal{G} \) denotes the gender group of the corresponding user. Here, \( y_i = 1 \) indicates a genuine authentication attempt and \( y_i = 0 \) indicates an impostor attempt. Let \( f_{\theta}: \mathbb{R}^{d} \rightarrow \{0,1\} \) be a trained authentication model parameterized by \( \theta \), where \( f_{\theta}(\mathbf{x}) = 1 \) denotes acceptance and \( f_{\theta}(\mathbf{x}) = 0 \) denotes rejection.

For a gender group \( g \in \mathcal{G} \), the group-specific False Acceptance Rate and False Rejection Rate are defined as
\[
\mathrm{FAR}_{g}
=
\Pr \left( f_{\theta}(\mathbf{x}) = 1 \mid y = 0, g \right),
\]
and
\[
\mathrm{FRR}_{g}
=
\Pr \left( f_{\theta}(\mathbf{x}) = 0 \mid y = 1, g \right).
\]
A gender bias exists if the authentication error distributions differ systematically across gender groups. Therefore, for two gender groups \(g_a\) and \(g_b\), our evaluation tests whether
\[
H_0: \mathrm{FAR}_{g_a} \sim \mathrm{FAR}_{g_b}
\quad \text{and} \quad
\mathrm{FRR}_{g_a} \sim \mathrm{FRR}_{g_b},
\]
against the alternative hypothesis
\[
H_1: \mathrm{FAR}_{g_a} \not\sim \mathrm{FAR}_{g_b}
\quad \text{or} \quad
\mathrm{FRR}_{g_a} \not\sim \mathrm{FRR}_{g_b}.
\]
In this setting, failure to reject \(H_0\) indicates no statistically significant evidence of gender disparity in authentication errors, while rejection of \(H_0\) indicates potential demographic bias. Our goal is not only to measure overall authentication accuracy, but to determine whether comparable error behavior is maintained across gender groups.
 
\underline{\textit{Contributions}}
This paper makes the following contributions:
(i) it presents one of the first empirical evaluations of gender fairness in swipe-based authentication using two publicly available datasets;
(ii) it compares both classical machine learning (XGBoost) and deep learning (DenseNet) authentication models; and
(iii) it evaluates demographic fairness using complementary statistical tests on False Acceptance Rate and False Rejection Rate distributions.

\section{Related work}
Prior studies have demonstrated that biometric systems, particularly face recognition, can exhibit demographic disparities, raising concerns about fairness in real-world deployment \cite{buolamwini2018gender}. Swipe-based authentication has consistently shown promising authentication performance and usability \cite{TouchSurveyTechniques, serwadda2013verifiers, GANTouch2022, FairFair}, yet its fairness across demographic groups remains largely unexplored. Although recent work has reported preliminary gender fairness observations for touch-based authentication \cite{GANTouch2020}, independent evaluations on additional datasets and authentication models are still limited. This study addresses that gap by systematically analyzing gender differences in False Acceptance Rate (FAR) and False Rejection Rate (FRR) using two publicly available swipe authentication datasets.

\begin{figure*}[htp]
\centering
\begin{tabular}{cccc}
% First row: BBMAS dataset
\begin{subfigure}{0.23\textwidth}
    \centering
    \includegraphics[width=\linewidth]{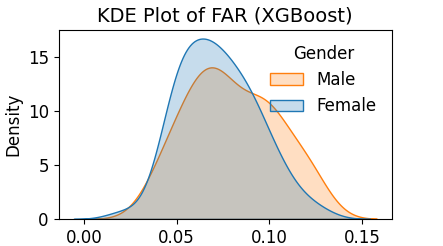}
    \caption{BBMAS FAR (XGB)}
    \label{BBMAS_XGB_FAR}
\end{subfigure} &
\begin{subfigure}{0.23\textwidth}
    \centering
    \includegraphics[width=\linewidth]{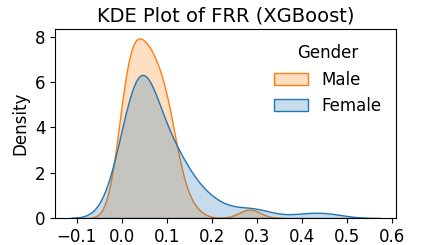}
    \caption{BBMAS FRR (XGB)}
    \label{BBMAS_XGB_FRR}
\end{subfigure} &
\begin{subfigure}{0.23\textwidth}
    \centering
    \includegraphics[width=\linewidth]{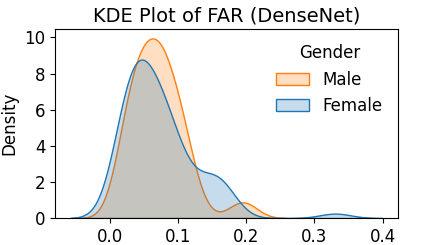}
    \caption{BBMAS FAR (DenseNet)}
    \label{BBMAS_DenseNet_FAR}
\end{subfigure} &
\begin{subfigure}{0.23\textwidth}
    \centering
    \includegraphics[width=\linewidth]{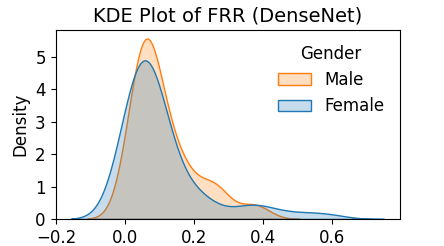}
    \caption{BBMAS FRR (DenseNet)}
    \label{BBMAS_DenseNet_FRR}
\end{subfigure} \\  % Move to the next row

% Second row: ANTAL dataset
\begin{subfigure}{0.23\textwidth}
    \centering
    \includegraphics[width=\linewidth]{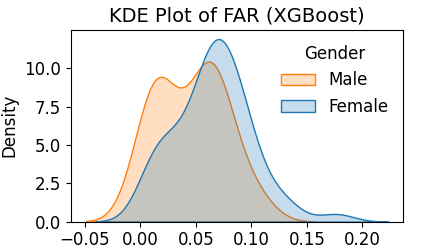}
    \caption{ANTAL FAR (XGB)}
    \label{ANTAL_XGB_FAR}
\end{subfigure} &
\begin{subfigure}{0.23\textwidth}
    \centering
    \includegraphics[width=\linewidth]{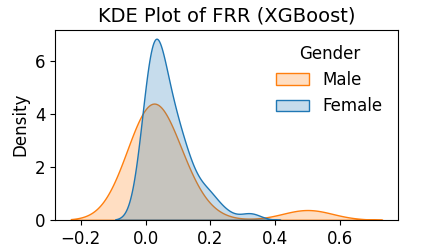}
    \caption{ANTAL FRR (XGB)}
    \label{ANTAL_XGB_FRR}
\end{subfigure} &
\begin{subfigure}{0.23\textwidth}
    \centering
    \includegraphics[width=\linewidth]{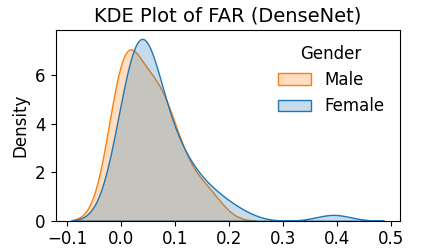}
    \caption{ANTAL FAR (DenseNet)}
    \label{ANTAL_DenseNet_FAR}
\end{subfigure} &
\begin{subfigure}{0.23\textwidth}
    \centering
    \includegraphics[width=\linewidth]{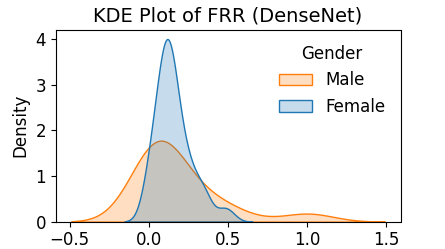}
    \caption{ANTAL FRR (DenseNet)}
    \label{ANTAL_DenseNet_FRR}
\end{subfigure}
\end{tabular}
\caption{Kernel density estimates of FAR and FRR for male and female users on the BBMAS and ANTAL datasets using XGBoost and DenseNet. The substantial overlap between distributions provides preliminary visual evidence of comparable authentication performance across genders, motivating the subsequent statistical analysis.}
\label{Combined_Results}
\end{figure*}

\section{Methodology} 
\label{MaterialMethods}
\underline{\textit{Datasets}}
In this study, we use two biometric authentication datasets, BBMAS \cite{belman2019insights} and ANTAL, which contain authentication data from users with varied male-to-female ratios, with 72/45 for BBMAS and 56/15 for ANTAL. The BBMAS dataset has 117 users, whereas ANTAL contains 71. These datasets provide a broad sample for evaluating authentication methods and investigating gender bias across gender groups. Unfortunately, most publicly available touch authentication datasets do not include demographic attributes \cite{FairFair}. 

\underline{\textit{Classification models}} We train our authentication models using two classifiers: XGBoost \cite{XGBoost} and DenseNet \cite{9423394}. XGBoost is a gradient boosting algorithm known for its efficiency and ability to handle structured data. In contrast, DenseNet is a deep learning architecture that increases feature propagation by forming dense connections between layers. We aim to compare the effectiveness of classical machine learning and deep learning models for biometric authentication and to investigate gender bias. 

\underline{\textit{Evaluation metric}} Lastly, we evaluate the effectiveness of these models using two essential metrics: false acceptance rate (FAR) and false rejection rate (FRR). FAR estimates the percentage of unauthorized users falsely accepted by the system, whereas FRR quantifies the percentage of genuine users incorrectly rejected.

\underline{\textit{Bias investigation process}} Let
\(
D_{m}=\{(\mathrm{FAR}_{i},\mathrm{FRR}_{i})\}_{i=1}^{n_{m}}
\)
and
\(
D_{f}=\{(\mathrm{FAR}_{j},\mathrm{FRR}_{j})\}_{j=1}^{n_{f}}
\)
denote the authentication error distributions for the male and female groups, respectively. Following the problem formulation, we investigate whether these distributions differ significantly across genders. For each classifier and dataset, we compare the FAR and FRR distributions using the Kolmogorov--Smirnov (KS), Mann--Whitney, and Wasserstein permutation tests.

The null hypothesis assumes that authentication errors are drawn from the same underlying distribution,
\[
H_{0}: D_{m} \equiv D_{f},
\]
indicating no measurable gender bias. The alternative hypothesis,
\[
H_{1}: D_{m} \neq D_{f},
\]
suggests statistically significant differences in authentication error distributions and therefore potential demographic bias. A significance level of
\(
\alpha=0.05
\)
is adopted throughout the study. If the resulting p-value satisfies
\(
p<\alpha,
\)
we reject \(H_{0}\); otherwise, we fail to reject \(H_{0}\), indicating no statistically significant evidence of gender disparity in authentication performance.

\section{Results and discussion}

Table \ref{tab:classificationresults} presents the accuracy, FAR, and FRR for both classifiers on the BBMAS and Antal datasets. XGBoost achieved authentication accuracies of 92\% on the BBMAS dataset and 94\% on the ANTAL dataset, compared with 91\% and 93\%, respectively, for DenseNet. XGBoost also maintained lower False Rejection Rates (FRR) of 8\% and 7\%, compared with 9\% and 16\% for DenseNet, while both classifiers exhibited comparable False Acceptance Rates (FAR) ranging from 6\% to 9\%.

To assess gender bias, we analyze the KDE plots in Figure~\ref{Combined_Results}, which depict the density distributions of FAR and FRR for male and female users. The overlapping distributions provide preliminary visual evidence of comparable authentication error distributions across genders, which is subsequently evaluated using formal statistical hypothesis testing. 

Statistical tests in Table 2 further validate this observation. The Kolmogorov-Smirnov (KS), Mann-Whitney, and Wasserstein permutation tests yield p-values above the threshold (\(\alpha = 0.05\)) in most cases, indicating no statistically significant disparity in error rates between male and female users. The only exception is the DenseNet FRR on the Antal dataset, which shows a marginally significant p-value, warranting further investigation.

Across two datasets, two authentication models, and three complementary statistical tests, we observe no consistent evidence of gender-dependent authentication errors, indicating comparable authentication performance across male and female users.

\begin{table}[h]
\centering
\caption{Authentication performance of XGBoost and DenseNet on the BBMAS and ANTAL datasets. The table reports classification accuracy together with False Acceptance Rate (FAR) and False Rejection Rate (FRR), where lower FAR and FRR indicate better authentication performance.}
\label{tab:classificationresults}
    \begin{tabular}{lccc|ccc}
        \toprule
        \textbf{Classifier} & \multicolumn{3}{c|}{\textbf{BBMAS dataset}} & \multicolumn{3}{c}{\textbf{Antal dataset}} \\
        & \textbf{Acc} & \textbf{FAR} & \textbf{FRR} & \textbf{Acc} & \textbf{FAR} & \textbf{FRR} \\
        \midrule
        XGBoost  & 92\% & 8\% & 8\% & 94\% & 6\% & 7\% \\
        DenseNet & 91\% & 9\% & 9\% & 93\% & 6\% & 16\% \\
        \bottomrule
    \end{tabular}
\end{table}

\begin{table}[h]
    \centering
    \caption{P-values from Kolmogorov--Smirnov (KS), Mann--Whitney, and Wasserstein permutation tests comparing male and female authentication error distributions. A significance level of $\alpha=0.05$ is used, with $p<0.05$ indicating a statistically significant gender difference.}
    \begin{tabular}{llcc|cc}
        \toprule
        \textbf{Classifier} & \textbf{Stat Test} & \multicolumn{2}{c|}{\textbf{BBMAS P-Val}} & \multicolumn{2}{c}{\textbf{Antal P-Val}} \\
        & & \textbf{FAR} & \textbf{FRR} & \textbf{FAR} & \textbf{FRR} \\
        \midrule
        XGBoost  & KS Statistic    & 0.092 & 0.151 & 0.126 & 0.241 \\
        XGBoost  & Mann-Whit    & 0.100 & 0.149 & 0.046 & 0.075 \\
        XGBoost  & Wasser-Perm    & 0.097 & 0.070 & 0.063 & 0.115 \\
        DenseNet & KS Statistic & 0.641 & 0.417 & 0.294 & 0.040 \\
        DenseNet & Mann-Whit & 0.687 & 0.279 & 0.312 & 0.422 \\
        DenseNet & Wasser-Perm & 0.474 & 0.380 & 0.635 & 0.053 \\
        \bottomrule
    \end{tabular}
\end{table}

\underline{\textit{Limitations}} The ANTAL dataset has a gender imbalance. This may reduce statistical power and make small disparities harder to detect. Therefore, non-significant results do not prove fairness. The analysis also relies on binary gender labels and lacks additional demographic attributes, limiting broader or intersectional fairness evaluation.

\section{Conclusion and future work}
Our study found no statistically significant gender bias in swipe-based authentication across two publicly available datasets and two authentication models. Consistent results from the Kolmogorov--Smirnov, Mann--Whitney, and Wasserstein permutation tests indicate that False Acceptance Rate (FAR) and False Rejection Rate (FRR) are comparable between male and female users. These findings suggest that swipe-based authentication can achieve equitable performance while maintaining high authentication accuracy, supporting its potential as a trustworthy behavioral biometric modality. Future work will extend this analysis to larger and more diverse datasets, additional demographic attributes, and other behavioral biometric modalities to further advance fairness-aware authentication systems.

\balance % to balance the references on both sides
\bibliographystyle{abbrv}
\bibliography{sigproc} 
\end{document}